# *Title: Analysis of the Second-Harmonic Generation Process in Nonlinear Metasurfaces Based on Multiple Quantum Wells: The Approach of Nonlinear Current Density, and The Development of Boundary Conditions*

***Authors**: **Sepehrdad Bahrami***
*School of Electrical and Computer Engineering, College of Engineering, University of Tehran, Tehran, Iran*

*E-mail: sepehr.bahrami@ut.ac.ir*

## *Abstract:*

*Using plasmonic metasurfaces integrated with nonlinear materials, such as multiple quantum wells (MQWs), has been an excellent solution to enhance nonlinear light-matter interactions. In this approach, a plasmonic resonator enhances the intensity of the fundamental harmonic within the MQWs region, increasing the strength of the intersubband transitions inside the MQWs. As a result, the efficiency of second harmonic generation (SHG) increases significantly.*

*In this paper, we present a new approach to analyze the SHG process in nonlinear metasurfaces based on MQWs. In this approach, first, we formulate the nonlinear current density within the MQW region, which serves as the radiation source of the second harmonic wave. Notably, the saturation effect in MQWs, that limits the nonlinear response, is accounted for. Second, we address the development of boundary conditions for the second harmonic (SH) field in periodic structures, which ultimately leads us to define periodic boundary conditions for the SH field. Third, we combine these elements into a semi-analytical framework and validate it against previously reported experimental data. Finally, an efficient MQWs structure and plasmonic metasurface are designed for the fundamental wavelength of* $8\ \mu m$. *After implementing the proposed algorithm, we observe that the designed nonlinear metasurface achieves a significant conversion efficiency of 0.02% under the incident intensity of* $23.5\ \frac{kW}{cm^2}$.

## *1. Introduction*

*Multiple-quantum-wells (MQWs) provide exceptionally large second-order nonlinear susceptibilities through resonantly enhanced intersubband transitions,*[1] *making them an attractive platform for highly efficient second harmonic generation (SHG) process. Unlike bulk nonlinear crystals, the subwavelength thickness of a MQW- based nonlinear metasurface substantially relaxes the longitudinal phase-matching constraint.*[2] *Nevertheless, conservation of in-plane momentum and the associated diffraction conditions remain important for determining the radiation channels of the generated harmonic.*

*In MQW-based nonlinear metasurfaces, the multiple-quantum-well region serves as the nonlinear element and consists of a one-dimensional periodic heterostructure. Within each period, semiconductor layers, including wells and barriers, are sequentially stacked, with different band gaps and nanometer-scale thicknesses. These heterostructures enable engineering of the electronic energy levels and, consequently, the intersubband optical transitions. A plasmonic resonator is typically integrated on the MQW substrate and designed such that its resonance*

*frequency coincides with the frequency of the incident wave, i.e., the fundamental frequency.*[3,4] *At resonance, the electric field component perpendicular to the MQW layers is strongly enhanced, thereby increasing the strength of the intersubband transitions and enhancing the second-harmonic generation (SHG) efficiency.*

*In literature* [3,5,6,4] *a few metasurfaces generating the second harmonic were realized and fabricated, by embedding MQWs in a plasmonic metasurface. In paper*[3]*, an experimental configuration for measuring the SH response was proposed for the first time. In article* [6]*, the SHG process in nonlinear metasurfaces has been analyzed and formulated for the first time. By using the reciprocity theorem, the effective second-order susceptibility is obtained, and then, the nonlinear response of the metasurface, such as conversion efficiency, has been calculated. In which saturation effect is taken into account. This formulation or approach has limitations regarding the incident angle, which must be normal to the structure.*

*In articles*[7,8] *nonlinear metasurfaces based on MQWs for third harmonic generation (THG) were realized and designed, and inspired by the approach in article*[6]*, the THG process has been formulated and analyzed by using the reciprocity theorem and finding effective third-order susceptibility. In paper*[8]*, the saturation factor for the THG process has been calculated for the first time, and additionally, two quantum wells were designed, one low-detuned and another one high-detuned, to study the detuning effect on conversion efficiency, and it was concluded that the same metasurface unit cell, along with a high-detuned MQWs, reaches saturation for the higher incident intensity and also achieves higher conversion efficiency.*

*In articles* [2,9] *nonlinear metasurfaces based on MQWs for the difference frequency generation (DFG) have been realized and designed. In paper* [2]*, the formulation for the analysis of the DGF process in these structures is presented. Additionally, the saturation effect in the calculation of the nonlinear response is accounted for. This research claims that this method of THz wave generation offers a few advantages over other methods. For instance: 1) the power of the THz wave is significant; 2) the generation of the THz wave via the DFG process is feasible at room temperature and without cooling.*

*For the first time, paper*[10] *has used the Pancharatnam-Berry (PB) phase approach to steer the far-field beam of the SH wave in nonlinear metasurfaces based on MQWs and has designed a metasurface for this purpose. The paper*[11] *has addressed the extension of the PB approach for nonlinear metasurfaces and has developed a theoretical framework, inspired by linear phase arrays. Ultimately, the designed metasurface enables control of the nonlinear beam. For instance, beam steering and vortex beam generation.*

*Article*[12] *proposed a nonlinear metasurface based on MQWs, in which, by applying different voltages to the MQWs of adjacent cells in one direction, the Stark effect is put into action, leading to the conduction band and MQWs parameters of adjacent cells having different slopes and values, respectively. In this case, the volume current, generating the SH wave, of each cell acquires a phase difference with respect to its adjacent cell. It is evident that in such an array with this description, the far-field beam of the SH wave will be steered.*

*The difference between the two approaches employed in recent papers to steer the nonlinear beam, PB and the Stark effect, is that the mechanism of creating a phase gradient is due to the rotation of the cell's geometry in the first approach, while in the second approach, it is due to applying voltage. Meaning that after fabrication, unlike the first approach, by employing the second approach, the nonlinear beam can be steered dynamically.*

*In article[13], the adjoint method is employed in topology optimization to inverse-design a nonlinear metasurface, whose effective second-order susceptibility is high. This algorithm is iterative and, after a few iterations, converges on a unit cell geometry with the highest conversion efficiency. Notably, the geometries obtained by this algorithm are asymmetric and unintuitive.*

*The theoretical description of MQW-based nonlinear metasurfaces has largely relied on effective nonlinear susceptibilities derived from reciprocity-based formulations. Such approaches provide an efficient description of the nonlinear response and incorporate saturation effects, but their formulation is naturally tied to specific excitation and radiation configurations. In particular, the treatment of arbitrary incidence and spatially distributed nonlinear sources in periodic structures is not explicit. A formulation that directly retains the spatial dependence of the nonlinear source would therefore provide a complementary and more general full-wave description of harmonic generation in these systems.*

*In this work, we develop such a formulation by representing the second-harmonic nonlinear polarization within the MQW region as an equivalent volume current density and coupling this source to a frequency-domain full-wave simulation at the second-harmonic frequency. The main contributions of this work are threefold. First, we formulate the second-harmonic source in MQW-based metasurfaces as a spatially distributed nonlinear volume current determined directly from the local fundamental electric field and the saturation-dependent second-order susceptibility. Second, we derive the Floquet phase relation required to impose periodic boundary conditions on the generated second-harmonic field in a periodic structure. Third, we combine these elements into a two-step full-wave computational framework and validate it against previously reported experimental data. The resulting framework provides a direct route from the spatially resolved fundamental field to the radiated harmonic power and conversion efficiency.*

*The rest of the paper is organized as follows. Section 2 presents an analytical expression for the intrinsic second-order susceptibility of multiple-quantum-well (MQW) structures. The effect of saturation on the nonlinear response is then considered, which actually restricts the second harmonic generation (SHG). We subsequently formulate the nonlinear response in terms of an equivalent volume current density generated within the MQW region by the fundamental-frequency field. This nonlinear current acts as the radiation source of the second-harmonic field, and an analytical expression for the current density is derived. In Section 3, the boundary conditions for the second-harmonic field in periodic structures are developed. Based on wave-vector conservation in the SHG process and the Rayleigh expansion of the generated field, we establish the pseudo-periodicity of the second-harmonic field and derive the corresponding Floquet wave vector. Section 4 presents the proposed semi-analytical framework for simulating the SHG process. The procedure for calculating the radiated second-harmonic power and conversion efficiency is described, and the framework is validated against previously reported experimental data. In Section 5, an MQW structure is designed for operation in the mid-infrared spectral range and optimized for SHG. Section 6 presents the design of the MQW-based nonlinear metasurface and describes the corresponding design procedure. Finally, Section 7 presents the calculated nonlinear response of the metasurface, including the SHG conversion efficiency, and discusses the resulting physical insights, followed by Conclusion in Section 8.*

# 2. Modeling of nonlinear current density

## 2.1 Second-order susceptibility for MQWs

*From quantum perspective, the second-order intrinsic susceptibility tensor* $\bar{\bar{\chi}}^{(2)}$ *can be obtained by perturbation theory.*[14] *In Multiple Quantum Wells, the dominant component of the tensor* $\bar{\bar{\chi}}^{(2)}$ *is associated with the electric field polarized normal to the quantum-well layers. Accordingly, all elements of this tensor are zero except* $\chi^{(2)}_{ZZZ}$, *which is obtained from the equation below:*[6,15]

$$\chi^{(2)}_{ZZZ} = S^{SHG}(I^{w1}_{MQW})N_e \frac{e^3 z_{21} z_{32} z_{31}}{\varepsilon_0 (E - E_{21} + i\hbar\gamma_{21})(2E - E_{31} + i\hbar\gamma_{31})} \quad (1)$$

*where* $S^{SHG}\left(I^{w1}_{MQW}\right)$ *is the saturation factor for the SHG process, which is a function of the intensity within the MQWs region,* $N_e$ *is the average density of electrons in one period* $\Lambda$ *of MQWs,* $e$ *is the electron charge,* $\varepsilon_0$ *is the vacuum permittivity, E is the energy of the incident photon and is equal to* $E = \hbar\omega_1$, $\omega_1$ *is the angular frequency of the first harmonic,* $E_{if} = E_i - E_f$, $z_{if}$, $\gamma_{if}$ *(i,f=1, 2, 3) are respectively transition energies, dipole moments, and linewidths for the transition from the initial energy level i to the final energy level f. The saturation factor can be written in terms of the population density of electronic levels as:*[6,15]

$$S^{SHG}\left(I^{\omega_1}_{MQW}\right) = \frac{N_1 - 2N_2 + N_3}{N_e} \quad (2)$$

*where* $N_i$ *(i=1,2,3) is the population density of electronic levels. Parameter* $N_e$ *is calculated as* $N_e = \frac{1}{\Lambda}\int_\Lambda n_e(z)dz$.[15] *In the following, we address the calculation of the saturation factor for SHG.*

## 2.2 Modeling saturation effect for SHG process

*By increasing the intensity of the incident wave, intersubband absorption depletes the first level of the MQWs,*[14] *thereby restricting the nonlinear process. This phenomenon is called the saturation effect. In fact, the saturation factor* $S^{SHG}$ *given in Eq. (2) is a scalar field in MQWs and takes values between zero and one. In other words, in the regions with low intensity, the saturation factor approaches one, meaning most of the electrons are in the first level (*$N_e \cong N_1$*) and in the regions with high intensity, the first level is depleted of carriers, and the saturation factor approaches zero.*[6]

*For calculating the carrier population density in each energy level, rate equations must be written in the steady-state to form the following system of equations with respect to* $N_1, N_2, N_3$[16,6,17]

$$\frac{\partial N_1}{\partial t} = -\frac{\alpha_{12}(\omega_1) I^{\omega_1}_{MQW}}{\hbar\omega_1} - \frac{\alpha_{13}(2\omega_1) I^{2\omega_1}_{MQW}}{\hbar 2\omega_1} + \frac{N_2 - N_{20}}{\tau_{21}} + \frac{N_3 - N_{30}}{\tau_{31}} \to 0 \quad (3)$$

$$\frac{\partial N_2}{\partial t} = -\frac{N_2 - N_{20}}{\tau_{21}} - \frac{\alpha_{23}(\omega_1) I^{\omega_1}_{MQW}}{\hbar\omega_1} + \frac{\alpha_{12}(\omega_1) I^{\omega_1}_{MQW}}{\hbar\omega_1} + \frac{N_3 - N_{30}}{\tau_{32}} \to 0 \quad (4)$$

*By using the charge conservation law, the third equation is written as*

$$N_1 + N_2 + N_3 = N_e \tag{5}$$

*In the equations above,* $N_{i0}$ *is the population density of i-th level under thermal equilibrium (before incidence). In the case that the Fermi energy is closer to the first level, the optical heating effect is neglected, and* $\forall i > 1\ N_{i0} \approx 0$ *is assumed.*[2,6] *Also* $N_i$ *is the population density of i-th level after the incidence.* $\tau_{if}$ *is the relaxation time for the transition from the initial level i to the final level f.*

*Moreover,* $I_{MQW}^{\omega_1}$ *is the intensity of the first harmonic wave within the MQWs region, and is calculated as* $I_{MQW}^{\omega_1}(\vec{r}) = \left|\vec{P}_{av}^{\omega_1}(\vec{r})\right|$ *, in which* $\vec{P}_{av}^{\omega_1}(\vec{r})$ *is the time-averaged Poynting vector at the first harmonic frequency. In Eq. (3),* $I_{MQW}^{2\omega_1}$ *is the intensity of the second harmonic wave within the MQWs region. Because of low conversion efficiency in nonlinear processes,* $I_{MQW}^{2\omega_1} \approx 0$ *is assumed.*[6]

$\alpha_{ij}$ *is the absorption coefficient for the transition from the initial level i to the final level j, and is expressed according to the formula below:*[6,15,18]

$$\alpha_{ij}(\omega_p) = \frac{N_i - N_j}{N_e}\,\alpha_{ij}^{(0)}(\omega_p) = \frac{4(N_i - N_j)\omega_p\left(e z_{ji}\right)^2 \hbar\gamma_{ji}}{\varepsilon_0 n_{MQW}^{\omega_p} c\,\left[\left(E - E_{ji}\right)^2 + (\hbar\gamma_{ji})^2\right]} \tag{6}$$

$n_{MQW}^{\omega_p}$ *is the refractive index of MQWs at the frequency of* $\omega_p$*. Saturation intensity for different transitions is equal to* $I_{S_{ij}}^{\omega_p} = \frac{N_e\hbar\omega_p}{2\alpha_{ij}^{(0)}(\omega_p)\,\tau_{ji}}$*.*[6,17] *Finally, by solving the system of equations in Eqs. (3) -(5), the* $(N_1, N_2, N_3)$ *will be obtained, and by the substitution of them into Eq. (2), the saturation factor for SHG is obtained as:*[6]

$$S^{SHG}\left(I_{MQW}^{\omega_1}\right) = \frac{2I_{MQW}^{\omega_1}\left[I_{S_{12}}^{\omega_1}\left(1 + \frac{\tau_{21}}{\tau_{31}}\right) - I_{S_{23}}^{\omega_1}\left(1 + \frac{\tau_{32}}{\tau_{31}}\right)\right] + 4I_{S_{12}}^{\omega_1} I_{S_{23}}^{\omega_1}\left(1 + \frac{\tau_{32}}{\tau_{31}}\right)}{3(I_{MQW}^{\omega_1})^2 + 2I_{MQW}^{\omega_1}\left[I_{S_{12}}^{\omega_1}\left(1 + \frac{\tau_{21}}{\tau_{31}}\right) + 2I_{S_{23}}^{\omega_1}\left(1 + \frac{\tau_{32}}{\tau_{31}}\right)\right] + 4I_{S_{12}}^{\omega_1} I_{S_{23}}^{\omega_1}\left(1 + \frac{\tau_{32}}{\tau_{31}}\right)} \tag{7}$$

## 2.3 The source of SH wave radiation

*Assume that a time-harmonic wave with a frequency of* $\omega_1$ *is incident upon the metasurface:*

$$\vec{E}_{inc}(\vec{r}, t) = Re\{\vec{E}_{inc}^{\omega_1}(\vec{r})e^{j\omega_1 t}\} \tag{8}$$

*where* $\vec{E}_{inc}^{\omega_1}(\vec{r})$ *is the phasor of the incident electric field that can have any spatial profile (such as a Gaussian beam, a plane wave, etc.), and with no limit on the incidence angle. The incident electric field creates an electric field within the MQWs region:*

$$\vec{E}_{MQW}(\vec{r}, t) = Re\{\vec{E}^{\omega_1}(\vec{r})e^{j\omega_1 t}\} \tag{9}$$

$\vec{E}^{\omega_1}$ *is the phasor of the electric field within MQWs, which has three vector components* $\vec{E}^{\omega_1}(\vec{r}) = E_x^{\omega_1}(\vec{r})\,\vec{a}_x + E_y^{\omega_1}(\vec{r})\,\vec{a}_y + E_z^{\omega_1}(\vec{r})\,\vec{a}_z$ *. The second harmonic polarization density within the MQWs region is related to the first harmonic electric field components inside the same region and to the second-order intrinsic susceptibility of MQWs. In the case of a dispersive and anisotropic medium:*[6],[19],[14]

$$P_i^{2\omega_1}(\vec{r}) = \varepsilon_0 \sum_j \sum_k \chi_{ijk}^{(2)}(2\omega_1, \omega_1, \omega_1; \vec{r})\, E_j^{\omega_1}(\vec{r})\, E_k^{\omega_1}(\vec{r}) \quad \vec{r} \in MQW's\ domain \tag{10}$$

*The equation above is expressed in the frequency domain, where* $i, j, k = \{x, y, z\}$*, and the reason why the susceptibility exhibits spatial dependence is that it depends on the intensity within MQWs, which in turn is a function of position. Additionally, we know that all elements of the tensor* $\bar{\bar{\chi}}^{(2)}(\vec{r})$ *are zero except* $\chi_{zzz}^{(2)}(\vec{r})$*, which is associated with the electric field component perpendicular to the structure. Therefore, we arrive at the equation below:*[6]

$$P_z^{2\omega_1}(\vec{r}) = \varepsilon_0\, \chi_{zzz}^{(2)}(\vec{r})\, [E_z^{\omega_1}(\vec{r})]^2 \quad \vec{r} \in MQW's\ domain \tag{11}$$

*where* $\chi_{zzz}^{(2)}(\vec{r})$ *is a short form of* $\chi_{zzz}^{(2)}(2\omega_1, \omega_1, \omega_1; \vec{r})$*. The relationship between the volume current density and the polarization density can be expressed as* $\vec{J} = \frac{\partial \vec{P}}{\partial t}$ *and* $\vec{J} = j\omega\vec{P}$ *in time- and frequency-domain, respectively. Volume current density, as the source of the SH wave radiation, is obtained in the form below, that only has the z component*[6]

$$J_z^{2\omega_1}(\vec{r}) = j2\omega_1\varepsilon_0\, \chi_{zzz}^{(2)}(\vec{r})\, \left[E_z^{\omega_1}(\vec{r})\right]^2 \quad \vec{r} \in MQW's\ domain \tag{12}$$

## 3. *Developing the boundary conditions for SH wave in the nonlinear periodic structures*

### 3.1 *Bloch-Floquet theory and Rayleigh expansion for fundamental harmonic electromagnetic fields in two-dimensional structures*

*For a rectangular-lattice periodic structure, the fields at opposite boundaries of the unit cell satisfy the Bloch–Floquet boundary condition given by:*[20]

$$\vec{E}_{1,dst} = \vec{E}_{1,src} e^{-j\vec{k}_{F1}.(\vec{r}_{dst} - \vec{r}_{src})} \tag{13}$$

$\vec{E}_1$ , $\vec{k}_{F_1}$ *are the total electric field and the Floquet wave vector of the fundamental harmonic, respectively. Where, for a unit cell lying in the xy-plane,* $\vec{k}_{F_1} = k_{F_{1x}}\,\vec{a}_x + k_{F_{1y}}\,\vec{a}_y = \vec{k}_{inc}^{\parallel}$ *, here* $\vec{k}_{inc}^{\parallel}$ *is the tangential component of the incident wave vector. Assume that the diffractive structure has spatial periodicity of* $P_x$ and $P_y$ *in x and y directions, respectively. Mathematically, functions satisfying the Bloch–Floquet boundary condition, known as pseudo-periodic functions, can be expressed as the product of a periodic function and an exponential function:*[20]

$$\vec{E}_1(\vec{r}) = \vec{F}_1(\vec{r})\, e^{-j\vec{k}_{F1}.\vec{r}} \qquad \vec{r} = x\,\vec{a}_x + y\,\vec{a}_y \tag{14}$$

*Periodic vector function* $\overrightarrow{F_1}(\vec{r}) = \sum_{m=-\infty}^{\infty}\sum_{n=-\infty}^{\infty}\vec{u}_{mn}(z)\, e^{-j\left(m\frac{2\pi}{P_x}\vec{a}_x+n\frac{2\pi}{P_y}\vec{a}_y\right).\vec{r}}$ *is expanded into a 2D Fourier series, then is substituted into Eq. (14), and following a simple mathematical manipulation, we have:*[20]

$$\vec{E}_1(\vec{r}) = \sum_{m=-\infty}^{\infty}\sum_{n=-\infty}^{\infty}\vec{u}_{mn}(z)\, e^{-j\left(\vec{k}_{F1}+m\frac{2\pi}{P_x}\vec{a}_x+n\frac{2\pi}{P_y}\vec{a}_y\right).\vec{r}} \tag{15}$$

*Substituting the field expansion,* $\vec{E}_1(\vec{r})$, *into the homogeneous Helmholtz equation shows that, in homogeneous regions, each vectorial Fourier coefficient,* $\vec{u}_{mn}(z)$, *can be separated into a constant vector and a scalar function of coordinate z representing forward- and backward-propagating waves.*[20]

$$\vec{E}_1(\vec{r}) = \sum_{m=-\infty}^{\infty}\sum_{n=-\infty}^{\infty}\vec{E}_{01}\, e^{-j\left(\vec{k}_{F1}+m\frac{2\pi}{P_x}\vec{a}_x+n\frac{2\pi}{P_y}\vec{a}_y\right).\vec{r}}\, [A_{mn}e^{-jk_{1z}z} + B_{mn}e^{jk_{1z}z}] \tag{16}$$

*Ultimately, in a homogeneous medium, the first harmonic electric field, which is a sum of all diffraction orders and the incident wave, can be written in terms of the Rayleigh expansion.*[20]

$$\vec{E}_1(\vec{r}) = \sum_{m=-\infty}^{\infty}\sum_{n=-\infty}^{\infty}\vec{E}_{01}\, e^{-j\vec{k}_1^{\parallel}.\vec{r}}\, [A_{mn}e^{-jk_1^{\perp}z} + B_{mn}e^{jk_1^{\perp}z}] \tag{17}$$

$$\begin{cases} \vec{k}_1^{\parallel} = \vec{k}_{F1} + m\dfrac{2\pi}{P_x}\,\vec{a}_x + n\dfrac{2\pi}{P_y}\,\vec{a}_y \\ k_1^{\perp} = \sqrt{k_1^2 - \left|\vec{k}_1^{\parallel}\right|^2} \end{cases} \tag{18}$$

*Here* $\vec{k}_1^{\parallel}$, $k_1^{\perp}$ *are the tangential and normal components of the first harmonic diffraction orders wave vector, respectively.* $k_1$ *is the wavenumber of the medium at the first harmonic frequency. The Rayleigh solution implies that the total electromagnetic field can be expressed as a series of plane waves with discrete wave vectors. If the whole media consists of several homogeneous media, a Rayleigh expansion is written for each homogeneous region. Note that if a specific diffraction order with the number of* $(m', n')$ *is propagating in one region, it can be evanescent in another region.*

## *3.2 Review the wave vector conservation law in the SHG process*

*The SHG process is a degenerate case of three-wave mixing. The conservation law of photon energy and the conservation law of wave vector in the SHG process can be represented as follows:*[19]

$$\begin{cases} \hbar\omega_2 = 2\hbar\omega_1 \\ \vec{k}_2 = 2\vec{k}_1 \end{cases} \tag{19}$$

*Here* $\hbar\omega_1$, $\hbar\omega_2$ *are the photon energies of the first harmonic and the second harmonic, respectively,* $\vec{k}_1$, $\vec{k}_2$ *are the wave vectors of the first harmonic and the second harmonic, demonstrated in Figure 1 a and b.*

*As we observed in Section 3.1, in diffractive structures, the wave vector that describes all first harmonic diffraction orders is presented in Eq. (18), in which* $\vec{k}_{F_1} = \vec{k}_{inc}^{\parallel}$.[20]

*In the general case, wave vector conservation in the SHG process always holds unless proven otherwise. For this reason, we examine the conservation of the tangential and normal components of the wave vector separately.*

$$\vec{k}_2 = 2\vec{k}_1 \;\; \rightarrow \begin{cases} \vec{k}_2^{\parallel} \stackrel{?}{=} 2\vec{k}_1^{\parallel} \\ k_2^{\perp} \stackrel{?}{=} 2k_1^{\perp} \end{cases} \tag{20}$$

*First, the wave vector conservation for the tangential component is investigated.*

$$\vec{k}_2^{\parallel} = 2\vec{k}_1^{\parallel} = 2(\vec{k}_{inc}^{\parallel} + m\frac{2\pi}{P_x}\,\vec{a}_x + n\frac{2\pi}{P_y}\,\vec{a}_y) \tag{21}$$

$$\vec{k}_{inc}^{\parallel} = k_0^{FH}\, sin\,\theta\,(cos\,\varphi\;\vec{a}_x + \;sin\,\varphi\;\vec{a}_y) \tag{22}$$

*where $\theta$ and $\varphi$ are the angles of the incident wave vector, demonstrated in Figure1 c, and $k_0^{FH}$ is the vacuum wavenumber at the first harmonic frequency. As seen in Eq. (21), wave vector conservation holds for the tangential components, since all parameters, such as the period of the structure, the incident wavelength, and angles, are always constant and well-defined values.*

*Now, the wave vector conservation for the normal component is investigated. In what follows, we prove that the normal component of the wave vector is not conserved. In other words, $k_2^{\perp} \neq 2k_1^{\perp}$.*

*We know that the normal component of the first harmonic wave vector $k_1^{\perp}$ is obtained by the dispersion relation of the medium:*

$$k_1^{\perp} = \sqrt{k_1^2 - \left|\vec{k}_1^{\parallel}\right|^2} = \sqrt{[k_0^{FH} n(\omega_1)]^2 - \left|\vec{k}_1^{\parallel}\right|^2} \tag{23}$$

*where $n(\omega_1)$is the medium refractive index at the first harmonic frequency. Additionally, the normal component of the second harmonic wave vector $k_2^{\perp}$ is obtained by the dispersion relation of the medium, where $k_2$ , $k_0^{SH}$ are the medium and vacuum wavenumbers at the second-harmonic frequency, respectively.*

$$k_2^{\perp} = \sqrt{k_2^2 - \left|\vec{k}_2^{\parallel}\right|^2} = \sqrt{[k_0^{SH} n(2\omega_1)]^2 - \left|2\vec{k}_1^{\parallel}\right|^2} \tag{24}$$

*Because of wave vector conservation for the tangential components, in the equation above $\vec{k}_2^{\parallel} = 2\vec{k}_1^{\parallel}$ is substituted. Also, it is clear that $k_0^{SH} = 2k_0^{FH}$and by a simple substitution, we arrive at the equation below:*

$$k_2^{\perp} = 2\sqrt{[k_0^{FH} n(2\omega_1)]^2 - \left|\vec{k}_1^{\parallel}\right|^2} \tag{25}$$

*By examining relations (23) and (25), it is evident that if the refractive index of the medium at the first and the second harmonic frequencies is equal, i.e. $n(2\omega_1) = n(\omega_1)$, then the conservation of the normal component of the wave vector holds.*

*In the general case, the relation between the first and the second harmonic wave vectors is as follows, in which $\overrightarrow{\Delta k}$ is mismatch wave vector:*

$$\vec{k}_2 = 2\vec{k}_1 + \overrightarrow{\Delta k} \tag{26}$$

*As mentioned above, conservation holds for tangential components, meaning that* $\vec{k}_2^{\parallel} = 2\vec{k}_1^{\parallel}$ *, but conservation does not hold for the normal component in the general case, meaning that* $k_2^{\perp} \neq 2k_1^{\perp}$ *. Thus, the mismatch wave vector only has a normal component; in other words,* $\overrightarrow{\Delta k} = \Delta k^{\perp}\vec{a}_z$ *, and we also have the equations below:*

$$\overrightarrow{\Delta k} = \overrightarrow{\Delta k^{\parallel}} + \Delta k^{\perp}\vec{a}_z \quad \rightarrow \begin{cases} \overrightarrow{\Delta k^{\parallel}} = 0 \\ \Delta k^{\perp} = k_2^{\perp} - 2k_1^{\perp} \end{cases} \tag{27}$$

*In a nonlinear periodic structure in which second-harmonic photons are generated, the relation between the wave vectors of the first and the second harmonic is shown in Figure 1 d. This figure is a sample case for a specific diffraction number, and for all diffraction numbers, this relation holds.*

$$\begin{cases} \vec{k}_2^{\parallel} = 2\vec{k}_1^{\parallel} = 2(\vec{k}_{inc}^{\parallel} + m\dfrac{2\pi}{P_x}\,\vec{a}_x + n\dfrac{2\pi}{P_y}\,\vec{a}_y) \\ k_2^{\perp} = \sqrt{k_2^2 - \left|\vec{k}_2^{\parallel}\right|^2} \end{cases} \tag{28}$$

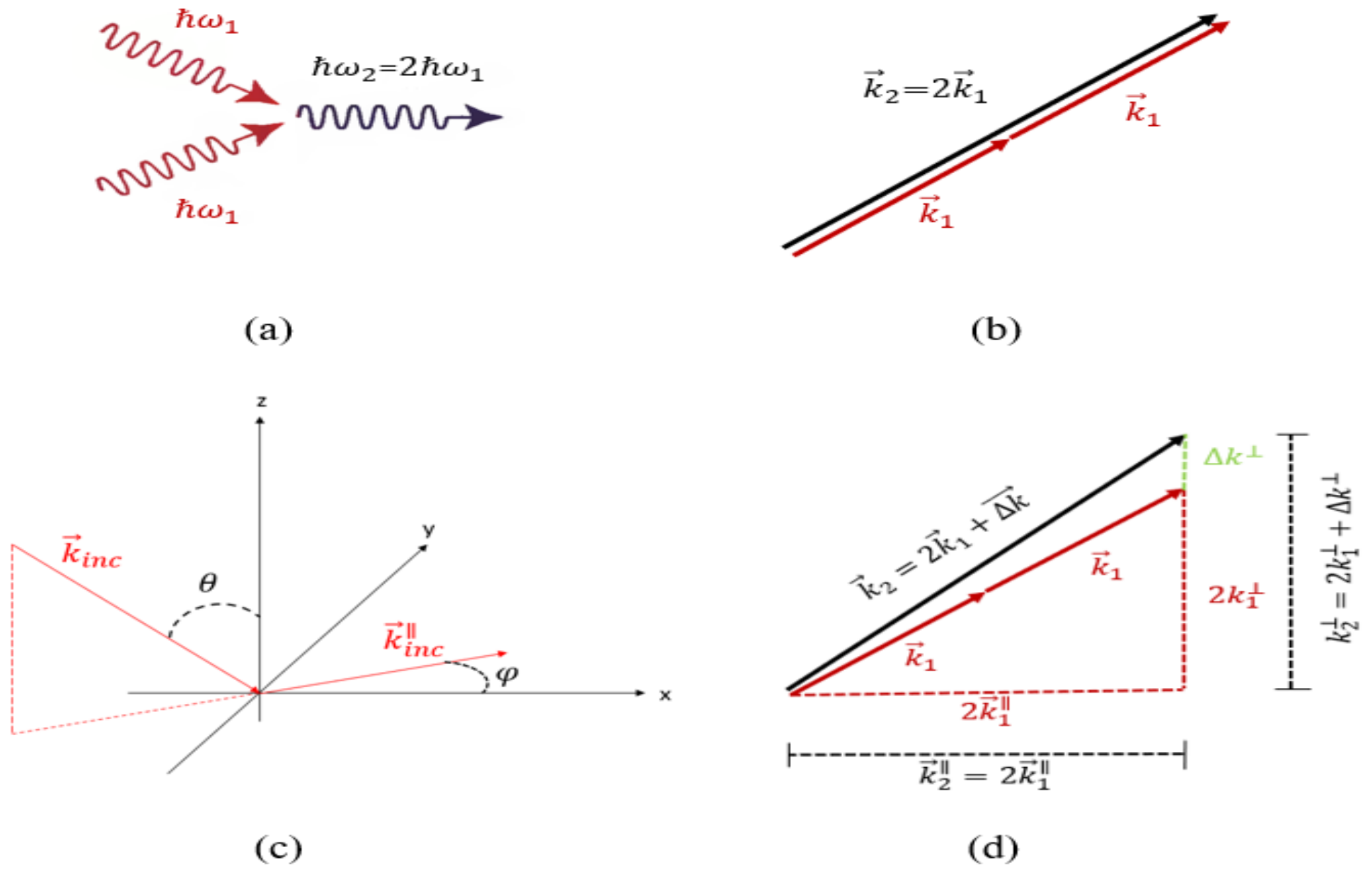

*Figure 1 . The conservation of (a) photon energy and (b) wave vector in the SHG process. (taken from[19]) (c) A view of the incident wave vector and the definition of incidence angles. (d) The relation between the wave vectors of the first harmonic and the second harmonic in the SHG process in nonlinear periodic structures.*

### 3.3 Using the Rayleigh solution and the wave vector conservation law in the SHG process to present a closed-form solution for the second harmonic electromagnetic field in diffractive structures

*In a second-order nonlinear periodic structure, the second harmonic total field, which is radiated by a volume current source, can be expanded in terms of Floquet modes (second harmonic diffraction orders), which is the Rayleigh solution itself, with the difference that the wave vector of second harmonic diffraction orders differs from that of the fundamental harmonic.*

$$\vec{E}_2(\vec{r}) = \sum_{m=-\infty}^{\infty}\sum_{n=-\infty}^{\infty} \vec{E}_{02}\, e^{-j\vec{k}_2^{\parallel}.\vec{r}}\, [C_{mn} e^{-jk_2^{\perp} z} + D_{mn} e^{jk_2^{\perp} z}] \tag{29}$$

*$\vec{E}_2(\vec{r})$ is the second harmonic electric field and parameters $\vec{k}_2^{\parallel}$, $k_2^{\perp}$ are tangential and normal components of the second harmonic wave vector and are obtained from Eq. (28). By the substitution of the existing expression for $\vec{k}_2^{\parallel}$ in Eq. (29), we have:*

$$\vec{E}_2(\vec{r}) = \sum_{m=-\infty}^{\infty}\sum_{n=-\infty}^{\infty} \vec{E}_{02}\, e^{-j(2\vec{k}_{F1}+2(m\frac{2\pi}{P_x}\vec{a}_x+n\frac{2\pi}{P_y}\vec{a}_y)).\vec{r}}\, [C_{mn} e^{-jk_2^{\perp} z} + D_{mn} e^{jk_2^{\perp} z}] \tag{30}$$

*In what follows, the product of a constant vector and a scalar function of z is written in the form of Fourier series coefficients.*

$$\vec{E}_2(\vec{r}) = \sum_{m=-\infty}^{\infty}\sum_{n=-\infty}^{\infty} \vec{v}_{mn}(z)\, e^{-j2(m\frac{2\pi}{P_x}\vec{a}_x+n\frac{2\pi}{P_y}\vec{a}_y).\vec{r}}\; e^{-j2\vec{k}_{F1}.\vec{r}} \tag{31}$$

$$\vec{E}_2(\vec{r}) = \sum_{m=-\infty}^{\infty}\sum_{n=-\infty}^{\infty} \vec{v}_{mn}(z)\, e^{-j(2m\frac{2\pi}{P_x}x+2n\frac{2\pi}{P_y}y)}\; e^{-j2\vec{k}_{F1}.\vec{r}} \tag{32}$$

*Due to the periodicity of the structure, the second harmonic electric field satisfies the Floquet boundary condition $\vec{E}_{2,dst} = \vec{E}_{2,src} e^{-j\vec{k}_{F2}.(\vec{r}_{dst}-\vec{r}_{src})}$. Therefore, a pseudo-periodic form exists for the second harmonic electric field:*

$$\vec{E}_2(\vec{r}) = \overrightarrow{\tilde{F}_2}(\vec{r})\, e^{-j\vec{k}_{F2}.\vec{r}} \tag{33}$$

*Where $\vec{k}_{F2}$ is the Floquet wave vector for the second harmonic. By examining the similarity of Eq. (32) to Eq. (33), we notice that:*

$$\vec{k}_{F_2} = 2\vec{k}_{F_1} \tag{34}$$

*Additionally, we conclude that $\overrightarrow{\tilde{F}_2}(\vec{r})$ is a periodic function with the period of $(P_x, P_y)$ and its Fourier series contains only even spatial harmonics.*

# 4. New approach for the simulation of the SHG process in nonlinear metasurfaces

*In this section, we address the simulation of the SHG process in nonlinear metasurfaces based on MQWs. For this purpose, we utilize COMSOL Multiphysics, in which two wave optics physics (frequency domain) are deployed: one for the first harmonic and another for the second harmonic simulation.*

*Review of the first physics configuration:*

*Initially, a unit cell of the metasurface is considered, and through a periodic port, a plane wave is incident upon the structure, where the direction of the incident wave vector $\vec{k}_{inc}$ can be expressed by the angles $\theta, \varphi$. Periodic boundary conditions are applied for opposite boundaries, where the Floquet wave vector of the first harmonic is $\vec{k}_{F_1} = \vec{k}_{inc}^{\parallel}$. Additionally, a Perfectly Matched Layer (PML) is placed behind the port. In this case, the reflected wave from the metasurface is absorbed by the PML and does not re-enter the computational domain to mimic the free space model. First wave optics physics is analyzed in the frequency domain at the first harmonic frequency $\omega = \omega_1$. Finally, electric and magnetic fields $\vec{E}^{\omega_1}(\vec{r})$, $\vec{H}^{\omega_1}(\vec{r})$ are obtained in the whole space.*

*Review of the second physics configuration:*

*Again, a unit cell of metasurface is considered, a volume current source $\vec{J}^{2\omega_1} = J_z^{2\omega_1}(\vec{r})\,\vec{a}_z$ is defined within the volume of MQWs, and periodic boundary conditions are applied for opposite boundaries, where the Floquet wave vector of the second harmonic is $\vec{k}_{F_2} = 2\vec{k}_{F_1}$. Additionally, to model radiation in free space, PML is used. Second wave optics physics is analyzed in the frequency domain at the second harmonic frequency $\omega = 2\omega_1$. Finally, electric and magnetic fields $\vec{E}^{2\omega_1}(\vec{r})$, $\vec{H}^{2\omega_1}(\vec{r})$ are obtained in the whole space.*

*Eventually, the algorithm for the simulation of the SHG process in nonlinear metasurfaces based on MQWs is proposed in Figure2. For better understanding, the simulation configurations of Physics 1 and Physics 2 are shown together in Figure3 a.*

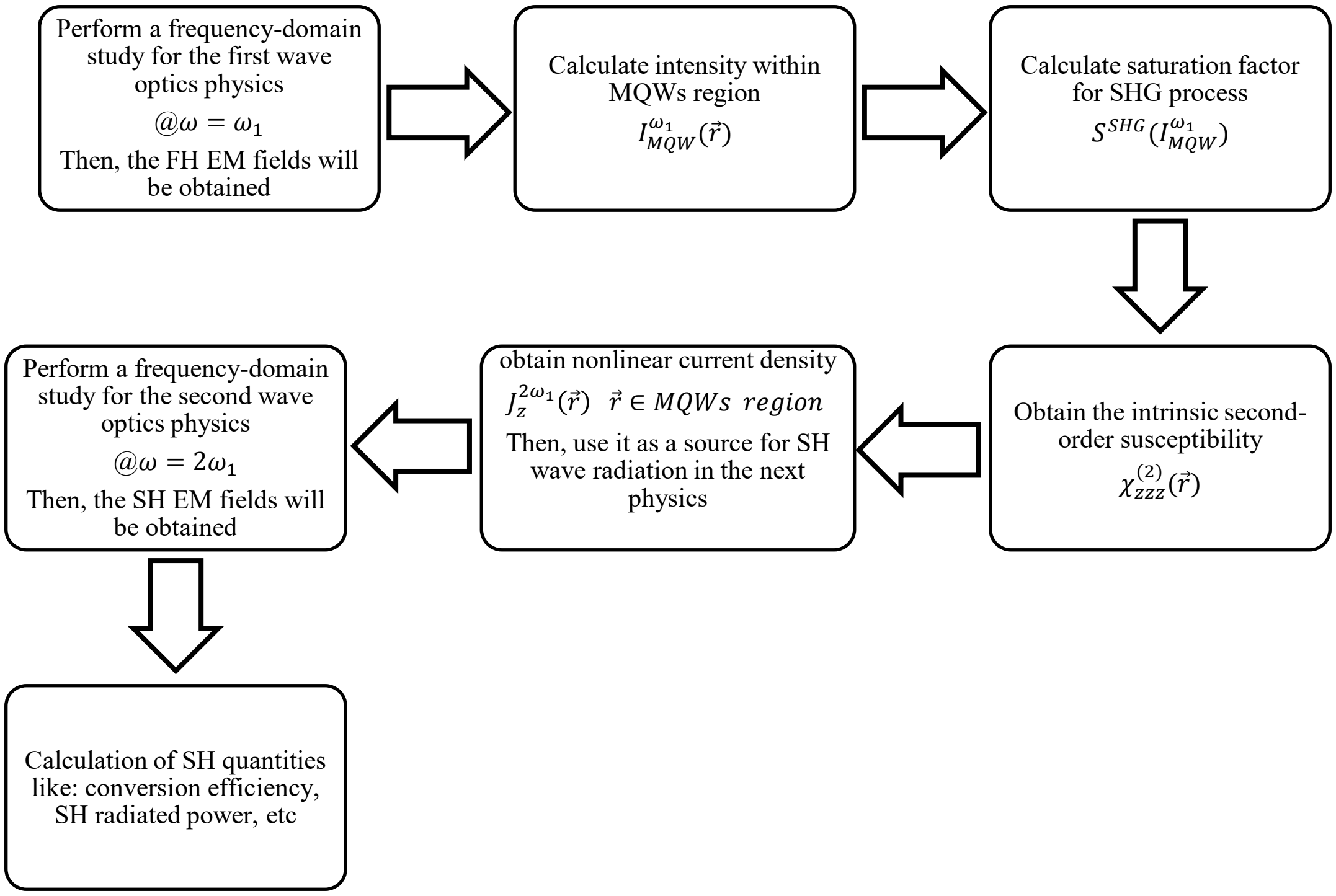

*Figure2 .The proposed algorithm for simulating the SHG process in nonlinear metasurfaces based on MQWs.*

## 4.1 Calculating the power of the generated SH wave and conversion efficiency

*After the simulation of the second physics, second harmonic electromagnetic fields are obtained. A surface called* $S_1$ *is considered in the far field, and the power of the generated SH wave,* $P^{SH}$*, is calculated through the equation* $P^{SH} = \iint_{S_1} \vec{P}_{av}^{2\omega_1}(\vec{r}).\hat{n}\, ds$ *, where* $\vec{P}_{av}^{2\omega_1}$ *is the time-averaged Poynting vector at the second harmonic frequency and* $\hat{n}$ *is the normal vector of the surface* $S_1$*.*

*As discussed in the appendix, if a plane wave is normally incident on a subwavelength periodic structure, the SH diffraction order with the number of* $(m,n) = (0,0)$ *is the only propagating one and travels in the perpendicular direction (z-direction). Thus,* $\vec{P}_{av}^{2\omega_1}(\vec{r})$ *only contains the z component:*

$$\vec{P}_{av}^{2\omega_1}(\vec{r}) = P_z^{2\omega_1}\vec{a}_z = \frac{1}{2}Re\left\{-E_y^{2\omega_1}\left(H_x^{2\omega_1}\right)^* + E_x^{2\omega_1}\left(H_y^{2\omega_1}\right)^*\right\}\vec{a}_z \quad (35)$$

*According to Eq. (35), the SH wave can be decomposed into two pairs of field components: the first pair,* $(-E_y, H_x)$ *and the second pair,* $(E_x, H_y)$*. Both pairs carry power in the z-direction. Based on the orientation of the electric-field component, the first and second pairs are referred to as y- and x- polarized components, respectively. This decomposition is motivated by conventional SHG measurements, in which the generated SH wave is typically passed through a polarizer before its power is measured by a detector.*[3] *Therefore, we separate the power of the generated SH wave for each polarization. By integrating each term on a surface in the far field, the power of the SH wave due to each polarization* $P_{y-polarized}^{SH}$ , $P_{x-polarized}^{SH}$ *is obtained.*

*In the SHG process, conversion efficiency is defined* $\eta = \frac{P^{SH}}{P^{FH}}$ *, here* $P^{FH}$ *is the power of the incident wave.*[6] *Using this equation, the total conversion efficiency (due to both polarizations) is obtained. We can also define conversion efficiency for the radiated SH wave with y or x polarization as* $\eta_{y-pol\ radiated\ wave} = \frac{P_{y-polarized}^{SH}}{P^{FH}}$ , $\eta_{x-pol\ radiated\ wave} = \frac{P_{x-polarized}^{SH}}{P^{FH}}$ .

## 4.2 Experimental Validation

*In this section, we intend to validate our proposed algorithm for simulating the SHG process. For this purpose, the paper*[3] *was selected, in which a nonlinear metasurface has been proposed. The wavelengths of the first and the second harmonics are* $\lambda_0^{FH} = 8\ \mu m$ *and* $\lambda_0^{SH} = 4\ \mu m$*, respectively. The first harmonic wave is y-polarized and normally incident upon the metasurface. After the SH wave is generated, it passes through a polarizer that transmits only y-polarization; its power is measured by a detector. Ultimately, the conversion efficiency graph is reported.*

*Using our proposed algorithm, the nonlinear metasurface was simulated in COMSOL Multiphysics, and the conversion efficiency for the radiated SH wave with y polarization was calculated. As observed in Figure3 b, simulation results obtained by the proposed algorithm closely matched the experimental results reported in the paper*[3]*.*

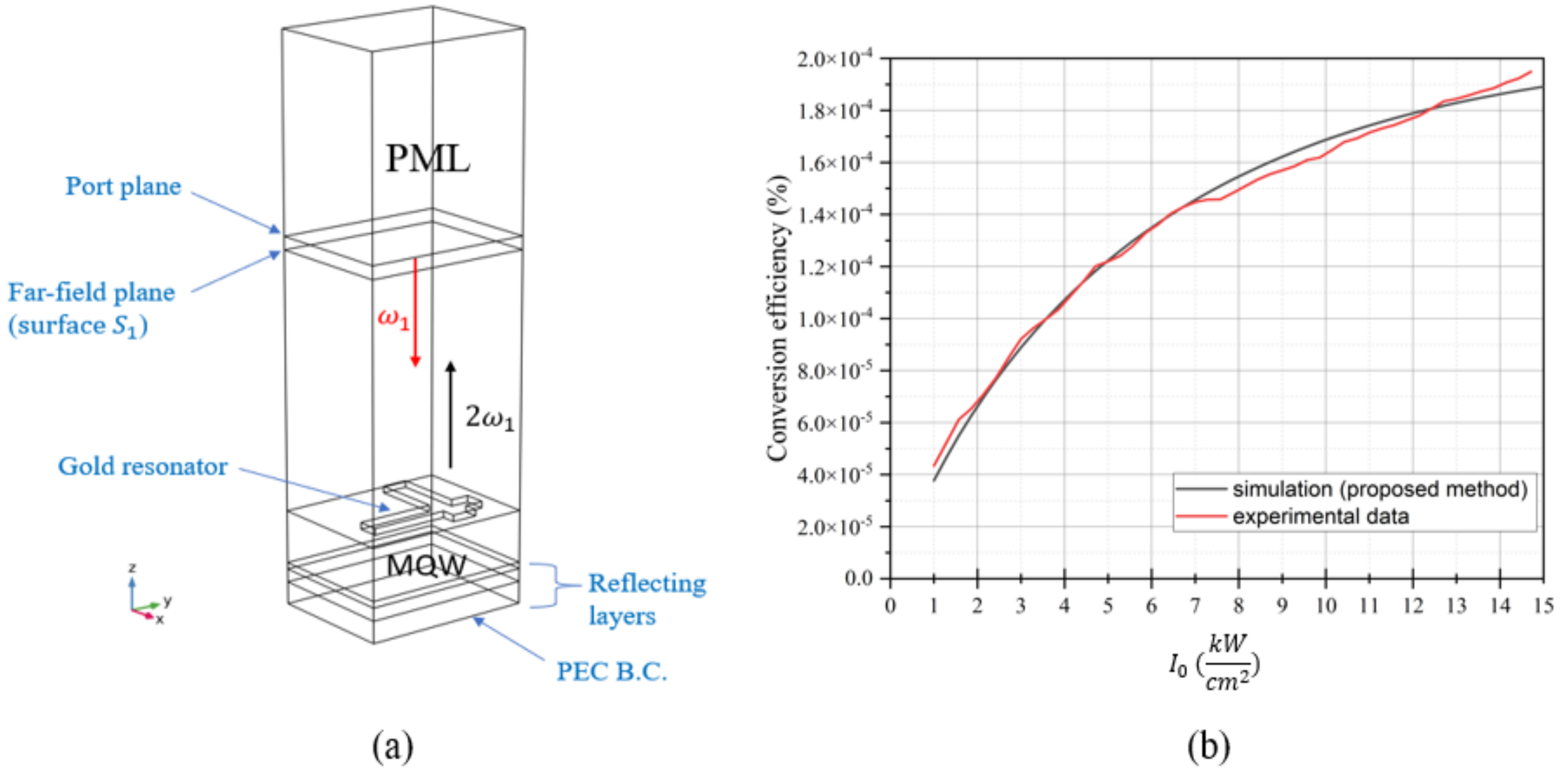

*Figure3. (a) The simulation configurations of Physics 1 and Physics 2, presented together in a single figure. (The geometry of nonlinear metasurface is taken from[3]) (b) Conversion efficiency in the SHG process calculated by the proposed approach (experimental data taken from[3]).*

## 5. Design of MQWs structure for SHG

*The structure of the designed MQWs consists of semiconductor materials $In_{0.53}Ga_{0.47}As/Al_{0.48}In_{0.52}As$, in which $In_{0.53}Ga_{0.47}As$ and $Al_{0.48}In_{0.52}As$ are the "well" and "barrier" materials, respectively. At this junction, the conduction band has a discontinuity of magnitude $\Delta E_c = 0.52\ eV$. To analyze and design the MQWs structure, a self-consistent Poisson-Schrödinger algorithm is employed, in which the non-parabolicity effects of the band are considered. This algorithm is implemented using LiveLink of COMSOL with MATLAB. Also, the electronic properties of both materials, such as electron effective mass ($m^*$), bandgap energy ($E_g$), static and high-frequency relative permittivities ($\varepsilon_s$, $\varepsilon_\infty$) and longitudinal-optical (LO) phonon energy ($\hbar\omega_{LO}$) are reported in Table1. Notably, in this table, $m_0$ is the mass of a free electron.[8]*

*The sequence of layers thickness in nm is $5/\mathbf{5.3}/1.3/\mathbf{2.2}/5$, where bold numbers indicate well layers. The first 2 nm of the first layer and the last 2 nm of the last layer are n-doped to a magnitude of $3 \times 10^{18}\ cm^{-3}$. After a few iterations, Poisson-Schrödinger algorithm is converged. Finally, the average electron density in one period is obtained $N_e = 5.9639 \times 10^{17}\ cm^{-3}$.*

*This MQWs structure is designed for the first and the second harmonic wavelengths equal to $\lambda_0^{FH} = 8\ \mu m$ and $\lambda_0^{SH} = 4\ \mu m$, respectively, so that it performs the SHG process optimally. It is worth mentioning that in this MQWs, the second level is detuned by $2.1\ meV$.*

*Table1. Electronic properties of semiconductor materials $Al_{0.48}In_{0.52}As$ and $In_{0.53}Ga_{0.47}As$ (taken from[8]).*

| $Material$ | $m^*$ | $E_g$ | $\varepsilon_s$ | $\varepsilon_\infty$ | $\hbar\omega_{LO}$ |
|---|---|---|---|---|---|
| $In_{0.53}Ga_{0.47}As$ | $0.041m_0$ | $0.75\ eV$ | 13.90 | 11.64 | $34\ meV$ |
| $Al_{0.48}In_{0.52}As$ | $0.071m_0$ | $1.46\ eV$ | 12.65 | — | — |

*In Figure4 a, potential energy along with probability density functions for different energy levels are demonstrated in one period. Additionally, in this figure, Fermi energy is shown. In the design process, an attempt has been made to keep the Fermi energy close to the first level so that, in this case, the optical heating effect is negligible.*

*After a series of calculations, the values of the designed MQWs parameters, such as transition energies $E_{ij}$, dipole moments $z_{ij}$, and relaxation times $\tau_{ij}$ are presented in Table2 . Notably, linewidth parameter $2\hbar\gamma_{ij}$ is not calculated due to its computational complexity; considering the measured values of this quantity in papers,[3,5] an approximate value is assumed.*

*Table2 . Calculated parameters of the designed MQWs.*

| **Transition** $i \rightarrow f$ | $E_{ij}$ (meV) | $z_{ij}$ (nm) | $\tau_{ij}$ (ps) | $2\hbar\gamma_{ij}$ (meV) (assumed) |
|---|---|---|---|---|
| 2 − 1 | *152.88* | *1.94* | *2.42* | ***19.2*** |
| 3 − 2 | *157.05* | *2.25* | *2.52* | ***20.55*** |
| 3 − 1 | *309.94* | *-0.68* | *2.76* | ***21.9*** |

*In Figure4 b, the absolute value of the second-order susceptibility versus wavelength is plotted for the designed MQWs. For simplicity of illustration, the saturation effect is not considered, meaning that the saturation factor is equal to one throughout the MQWs region, and in this case $\chi^{(2)}_{zzz}$ does not depend on position.*

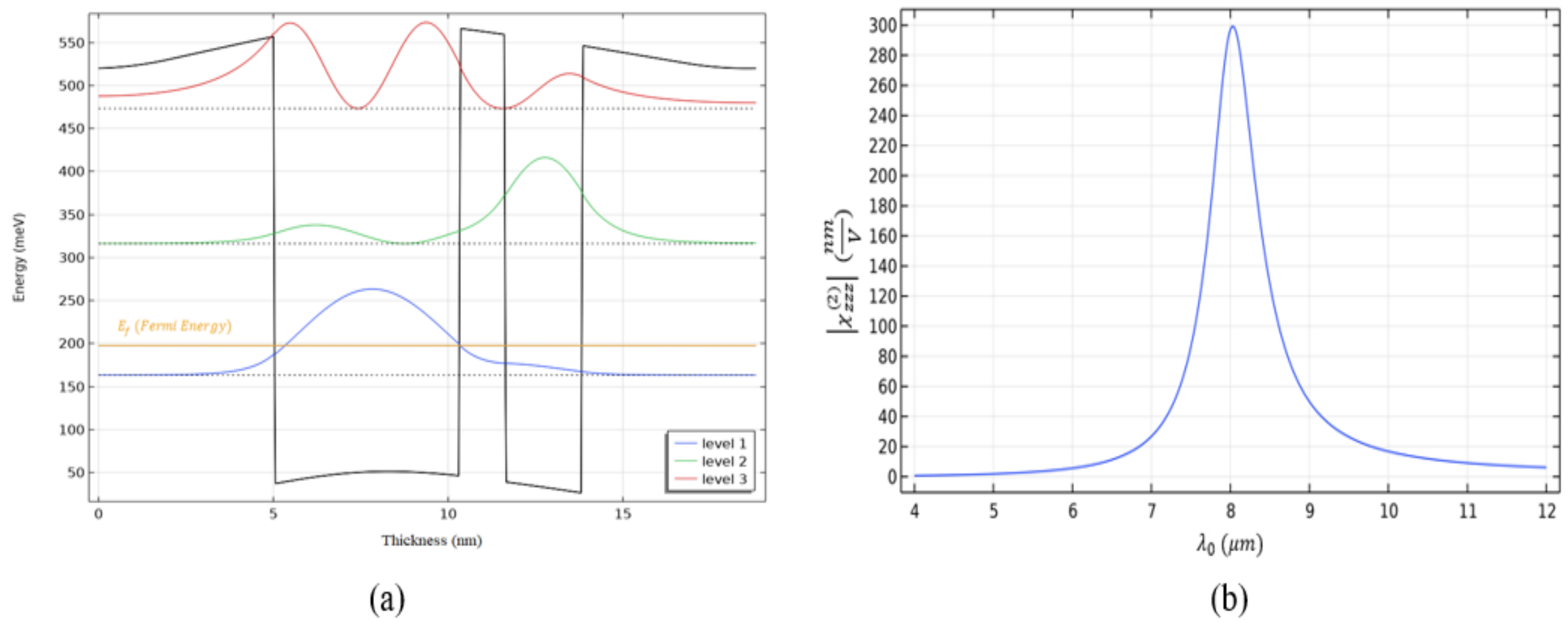

*Figure4 . (a) Potential energy diagram of MQW structure (solid black line) along with the probability density functions for each energy level (solid colored lines), eigenvalues of energy (black dashed lines) and Fermi energy (solid yellow line). (b) Diagram of $\left|\chi^{(2)}_{zzz}\right|$ versus wavelength without consideration of the saturation effect.*

## 6. Metasurface design

*As shown in Figure5 a and b, the proposed metasurface unit cell has a C-shaped geometry. In this unit cell, the ground plane is placed at the bottom of the metasurface to reflect the SH wave upward. The ground plane consists of 150 nm of gold, 50 nm of platinum, and 10 nm of titanium. Additionally, a gold resonator with a thickness of 50 nm is placed on top of the MQWs substrate, which has a thickness of 400 nm and consists of almost 21 periods.*

*The function of this resonator is to enhance the z-component of the FH electric field $E_z^{\omega_1}(\vec{r})$ within the MQW region so that the conversion efficiency for the SHG process increases. A plane wave at the wavelength of $\lambda_0^{FH} = 8\ \mu m$ is normally incident upon the metasurface and the generated SH wave at the wavelength of $\lambda_0^{SH} = 4\ \mu m$ propagates normally in free space. In the following, the second harmonic response is calculated under this excitation.*

*In the metasurface design, we performed a parametric sweep over the unit cell's geometric dimensions to find specific dimensions whose absorption spectrum has absorption peaks at the first and the second harmonic wavelengths. In Figure5 c, the absorption spectrum of the unit cell is observed for two different incident polarizations. According to Figure5 c, since there is an absorption peak for the incident polarization of x at the FH wavelength ($\lambda_0 = \lambda_0^{FH} = 8\ \mu m$). It is reasonable to choose an x-polarized plane wave incident upon the structure. In what follows, we calculate the nonlinear response for this incident polarization.*

*In Figure5 d, the normalized electric field is plotted for an x-polarized incident wave at the wavelength of $\lambda_0 = \lambda_0^{FH} = 8\ \mu m$. It is observed that the resonator produces an almost uniform distribution of $E_z^{\omega_1}$ throughout the MQWs.*

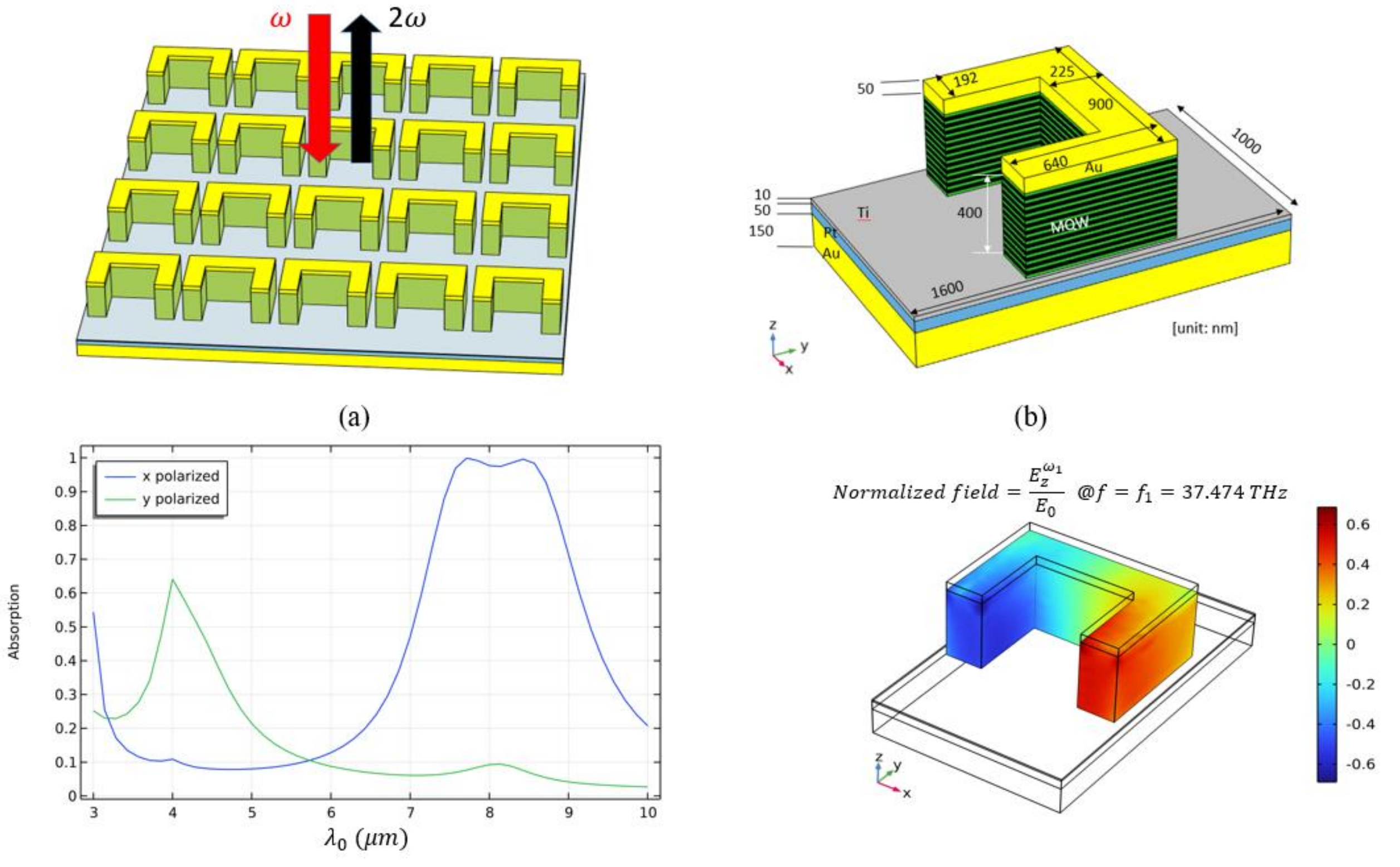

(a) (b)

(c) (d)

*Figure5 . (a) General scheme of nonlinear metasurface and the SHG process under normal incidence. (b) Details of designed unit cell. (c) Absorption spectrum of the designed metasurface for two incidence polarizations. (d) Normalized electric field (real part) for the designed metasurface, the incident plane wave is x-polarized at the wavelength of $\lambda_0 = \lambda_0^{FH} = 8\ \mu m$.*

## *7. Results and discussion*

*We employ our proposed algorithm, presented in Figure2 , to calculate the nonlinear response of the metasurface, including SH electromagnetic fields, and conversion efficiency. This algorithm is implemented in COMSOL Multiphysics. In this simulation, an x-polarized plane wave at the wavelength of $\lambda_0 = \lambda_0^{FH} = 8\ \mu m$ and of intensity $I_0 = 5\frac{kW}{cm^2}$ is incident on the structure.*

*In Figure6 a, the absolute value of the second-order susceptibility* $\left|\chi_{zzz}^{(2)}(\vec{r})\right|$ *versus position is plotted with the saturation effect considered. As observed, the value of* $\left|\chi_{zzz}^{(2)}(\vec{r})\right|$ *decreases near the resonator, since the intensity of the first harmonic wave is relatively high in the vicinity of the resonator, and consequently, the saturation effect is pronounced. In Figure6 b, the volume current density* $J_z^{2\omega_1}(\vec{r})$*, which generates the SH wave, is shown within the MQWs region.*

*From Figure5 d, it can be inferred that beneath the resonator's arms in the y-direction, from the top to the bottom of the MQWs, the field distribution of* $E_z^{\omega_1}$ *is uniform and has a significant value. Thus, according to the equation* $J_z^{2\omega_1}(\vec{r}) = j2\omega_1\varepsilon_0\,\chi_{zzz}^{(2)}(\vec{r})\,\left[E_z^{\omega_1}(\vec{r})\right]^2$ *it is evident that the volume current density within the aforementioned region is uniform. Meanwhile, in other papers,*[21],[3] *the designed unit cells have a significant field distribution of* $E_z^{\omega_1}$ *near the resonator, whereas elsewhere it approaches zero, leading to the resulting volume current density being mostly concentrated near the resonator and, consequently, the radiation of the SH wave by this source is weaker.*

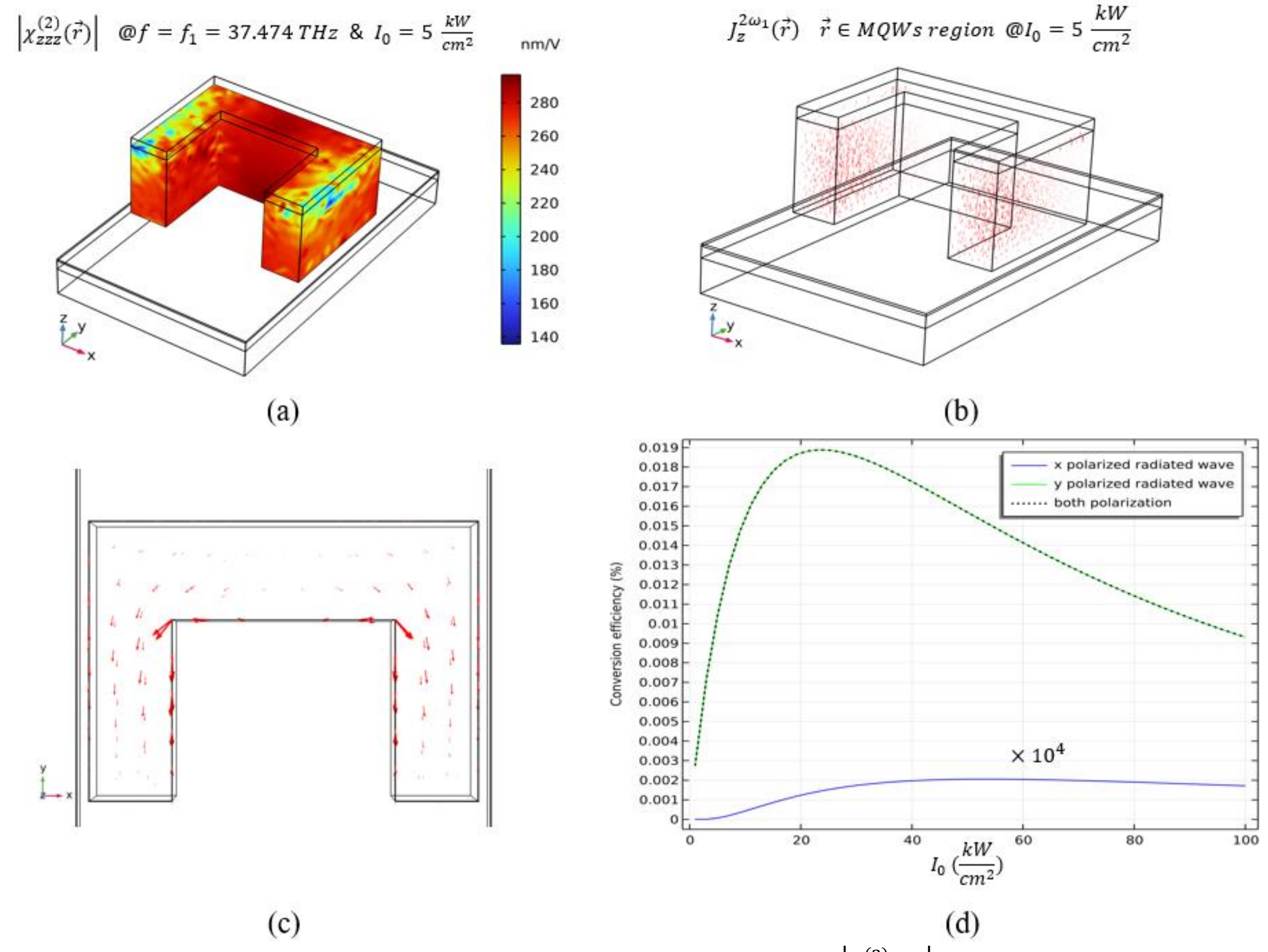

*Figure6 . (a) Spatial distribution of the second-order susceptibility magnitude,* $\left|\chi_{zzz}^{(2)}(\vec{r})\right|$ *, considering the saturation effect. (b) Volume current density* $J_z^{2\omega_1}(\vec{r})$ *(real part), as the excitation source of SH wave. (c) Surface current distribution (real part), induced on the gold surface by the volume current source. (d) Power conversion efficiency versus incident intensity for the SHG process. (Incident plane wave is x-polarized, and for better clarity, the blue plot is scaled by a factor of* $10^4$*).*

*Volume current density* $J_z^{2\omega_1}(\vec{r})$ *within the MQWs region as an excitation source, induces surface current on the gold at the wavelength* $\lambda_0 = \lambda_0^{SH}$*. In fact, gold acts as an antenna, resulting in stronger radiation. In Figure6 c, the* *induced surface current on the gold is plotted. In Figure 6d, conversion efficiency versus incident intensity $I_0$ is shown. Note that the conversion efficiencies for the radiated SH wave with x, y, and both polarizations are calculated separately. It is observed that as incident intensity increases, conversion efficiency initially increases, then reaches saturation and starts to decrease. The designed nonlinear metasurface achieves a significant conversion efficiency of* $0.02\ \%$ *under the incident intensity of* $I_0 = 23.5\ \frac{kW}{cm^2}$.

*From Kirchhoff's law of thermal radiation, we know that the absorption spectrum is equal to the emission spectrum under thermal equilibrium. Meaning that Figure 5c is the emission spectrum. When the unit cell operates in the emission mode ($\lambda_0 = \lambda_0^{SH} = 4\ \mu m$), the radiated SH wave contains both x and y polarization. Based on the emission spectrum diagram, it is expected that the emitted power with y polarization is much larger than that with x polarization. Since an appropriate surface current is induced on the gold, our expectation is met, and, ultimately, the SH wave with y-polarization has higher power and conversion efficiency.*

## 8. Conclusion

*In this research, a new approach was presented for analyzing the SHG process in nonlinear metasurfaces based on MQWs, in which the volume current within the MQWs serves as the radiation source of the second harmonic (SH) wave. We proposed that the electromagnetic field radiated by this volume current can be expanded in terms of SH diffraction orders. Thus, a closed-form solution was presented for the SH electromagnetic field. In the following, we defined periodic boundary conditions for the SH field, and an expression for its Floquet wave vector was obtained.*

*At the end of this approach, the calculation of SH power and conversion efficiency was discussed. Ultimately, this proposed approach was validated against previously reported experimental data. Briefly, while presenting a new approach, we succeeded in developing the mathematics related to the generation of SH wave in nonlinear periodic structures.*

*Then, an optimal MQWs structure and metasurface unit cell were designed for operation in the mid-IR spectral range, in which the SH wave ultimately propagates in free space. After implementing the proposed algorithm, it was observed that the designed nonlinear metasurface achieved a significant conversion efficiency of* $0.02\ \%$ *under the incident intensity of* $I_0 = 23.5\ \frac{kW}{cm^2}$.

*For future works, we can apply this proposed approach, which is based on the volume current within MQWs, to other nonlinear processes, including Third Harmonic Generation (THG), Difference Frequency Generation (DFG), and Sum Frequency Generation (SFG). Meanwhile, the mathematics related to the generation of nonlinear photons in the periodic structures can be developed.*

## Appendix: Specific case of normal incidence

*If the period of the structure is subwavelength, meaning that* $P_x, P_y < \lambda_1$, *where* $\lambda_1$ *is the medium wavelength at FH frequency, and also, the incidence is normal, meaning that* $\vec{k}_{inc}^{\parallel} = 0$, *or in other words,* $\vec{k}_{inc}^{\parallel}$ *has no component in either the x or y direction. In this case, the FH- and SH-diffraction orders with the number of* $(m, n) = (0,0)$ *are the only propagating ones. In what follows, the proof for this claim is presented:*

$$\vec{k}_1^{\parallel} = \vec{k}_{inc}^{\parallel} + m\frac{2\pi}{P_x}\,\vec{a}_x + n\frac{2\pi}{P_y}\,\vec{a}_y \qquad (36)$$

$$\vec{k}_1^{\parallel} = m\frac{2\pi}{P_x}\,\vec{a}_x + n\frac{2\pi}{P_y}\,\vec{a}_y \tag{37}$$

*The sufficient condition for the FH diffraction orders to be propagating is as follows:*

$$\left|\vec{k}_1^{\parallel}\right|^2 \le k_1^2 = \left(\frac{2\pi}{\lambda_1}\right)^2 \tag{38}$$

$$(m\frac{2\pi}{P_x})^2 + (n\frac{2\pi}{P_y})^2 \le \left(\frac{2\pi}{\lambda_1}\right)^2 \xrightarrow{yields} (m\frac{\lambda_1}{P_x})^2 + (n\frac{\lambda_1}{P_y})^2 \le 1 \tag{39}$$

*Since* $P_x\,, P_y < \lambda_1$, *as a result, we have* $\frac{\lambda_1}{P_x},\frac{\lambda_1}{P_y} > 1$, *therefore only in the case that* $\mathrm{m} = \mathrm{n} = 0$, *the inequality is satisfied. For the FH diffraction order with the number of* $(m,n) = (0{,}0)$, *which is the only propagating one, we have:*

$$\vec{k}_1^{\parallel} = \vec{k}_{inc}^{\parallel} + m\frac{2\pi}{P_x}\,\vec{a}_x + n\frac{2\pi}{P_y}\,\vec{a}_y \xrightarrow{\vec{k}_{inc}^{\parallel}=0\,,\ n=m=0} \vec{k}_1^{\parallel} = 0 \tag{40}$$

$$k_1^{\perp} = \sqrt{k_1^2 - \left|\vec{k}_1^{\parallel}\right|^2} \xrightarrow{\vec{k}_1^{\parallel}=0} k_1^{\perp} = k_1$$

*Thus, in the case that a plane wave is normally incident on the subwavelength 2D periodic structure, only the FH diffraction order with the number of* $(m,n) = (0{,}0)$ *is the propagating one and travels in the normal direction.*

*Now, under normal incidence, what would the SH electromagnetic fields look like?*

*We already know that the conversion law of the wave vector only holds for the tangential components:*

$$\vec{k}_2^{\parallel} = 2\vec{k}_1^{\parallel} \xrightarrow{\vec{k}_1^{\parallel}=0} \vec{k}_2^{\parallel}= 0 \tag{41}$$

*Since* $\vec{k}_1^{\parallel}$ *is zero, therefore* $\vec{k}_2^{\parallel}$ *is zero too, meaning that the wave vector of SH diffraction orders has no tangential components and propagates in the perpendicular direction, and the normal component of its wave vector is given by:*

$$k_2^{\perp} = \sqrt{k_2^2 - \left|\vec{k}_2^{\parallel}\right|^2} \xrightarrow{\vec{k}_2^{\parallel}=0} k_2^{\perp} = k_2 \tag{42}$$

*Therefore, in the case that a plane wave is normally incident on the subwavelength 2D periodic structure, only the SH diffraction order with the number* $(m,n) = (0{,}0)$ *is the propagating one and travels in the perpendicular direction. In Figure 7, the wave vectors of the first and second harmonics are shown under normal incidence on a subwavelength 2D periodic structure.*

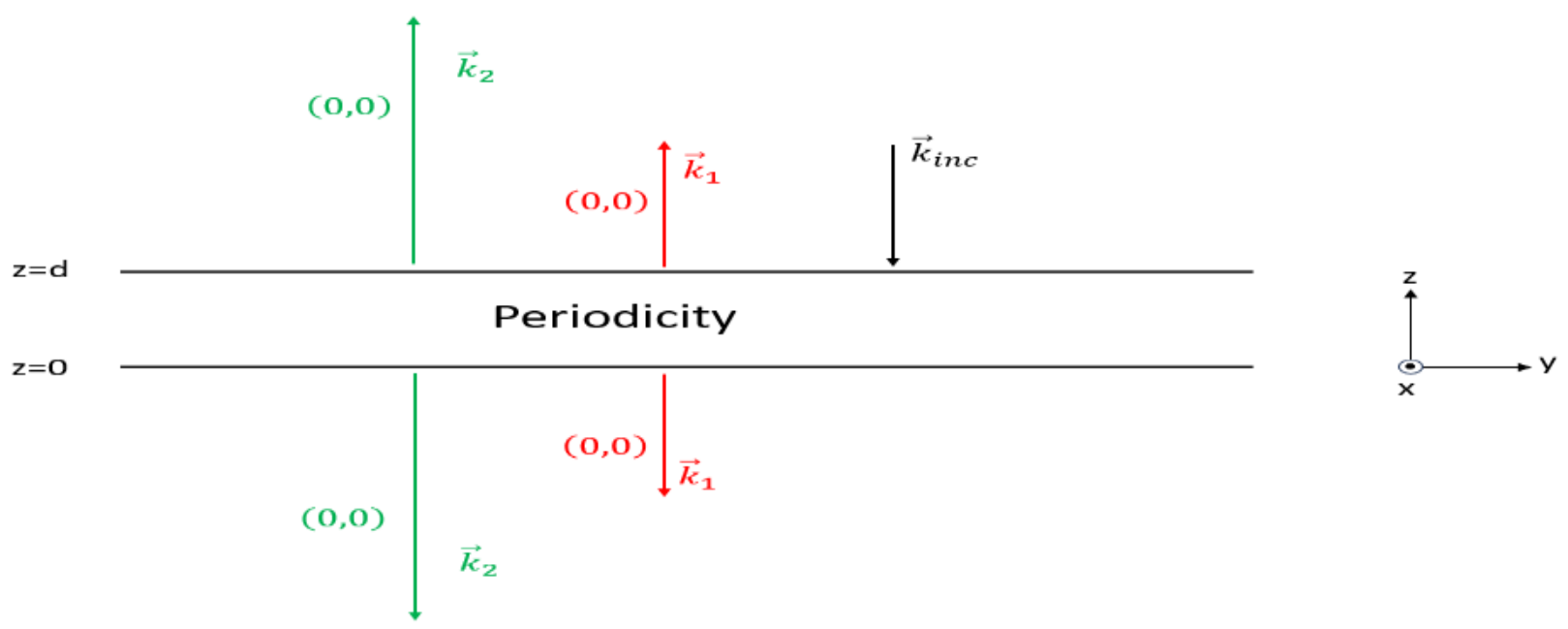

*Figure 7. Wave vectors of the first and the second harmonics in the case that the plane wave is normally incident on a subwavelength periodic structure.*

# *SUPPLEMENTARY INFORMATION*

## *Analysis of the Second-Harmonic Generation Process in Nonlinear Metasurfaces Based on Multiple Quantum Wells: The Approach of Nonlinear Current Density, and The Development of Boundary Conditions*

***Authors**: **Sepehrdad Bahrami***

*School of Electrical and Computer Engineering, College of Engineering, University of Tehran, Tehran, Iran*

***Abstract**:*

*For electromagnetic simulation, the permittivity and permeability of all materials must be specified. In this supplementary information, we first address the calculation of the permittivity of undoped Multiple Quantum Wells (MQWs), using effective medium theory. Then, we obtain the permittivity for doped MQWs.*

## *1. Permittivity tensor of Multiple Quantum Wells (MQWs)*

### *1.1 Calculation of the permittivity tensor of Undoped MQWs using effective medium theory*

*Multiple Quantum Wells (MQWs) is a one-dimensional periodic structure whose unit cell consists of a number of layers with different materials and thicknesses. As observed in Figure 1, layers are stacked on top of one another in the z direction. Notably, each layer is a homogeneous medium, and also can be dispersive.*

*The goal of effective medium theory is to obtain an equivalent permittivity for a one-dimensional periodic structure, which is mainly tensorial.*

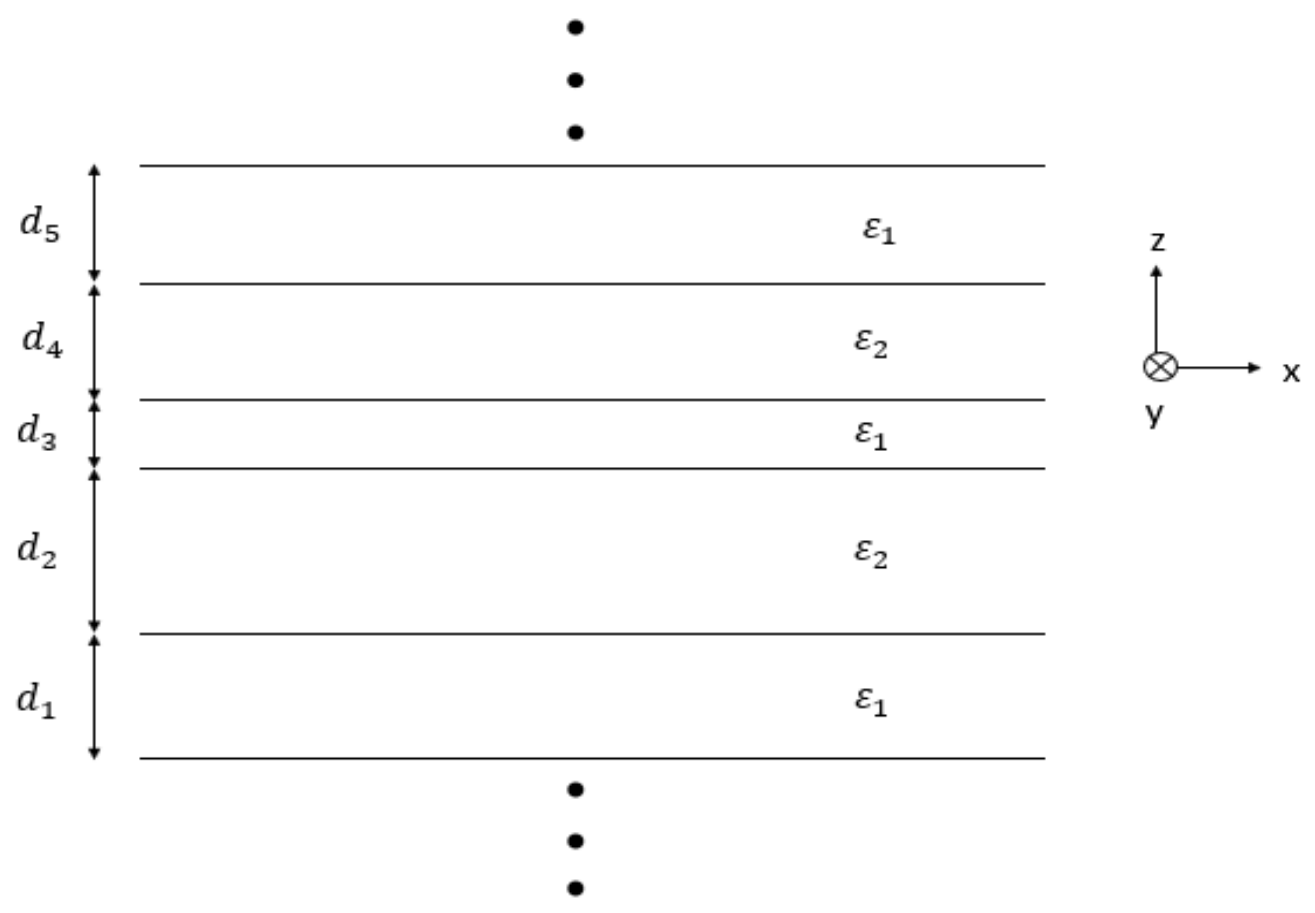

*Figure 1. One period of MQWs, which is 1D periodic structure. Permittivity and thickness of each layer are specified.*

*$\varepsilon_{core}(\omega)$ is the permittivity of MQW in the absence of doping:*[1]

$$\varepsilon_{core}(\omega) = \varepsilon_{MQW}(\omega)\big|_{Ne=0} \tag{1}$$

*Effective medium theory states that if the thicknesses of all layers are much smaller than the medium wavelength. (meaning $\forall i = 1, \ldots, M \quad d_i \ll \lambda$ ). In this case, the total field $\varphi = (\vec{E}, \vec{H}, \vec{D}, \vec{B})$ can be expressed as a weighted sum. $f_n$, $\varphi_n$ are the weight coefficient and the field of the medium n, respectively.*

$$\varphi = \sum_{n=1}^{M} f_n \varphi_n \qquad f_n = \frac{d_n}{\sum_{i=1}^{M} d_i} \tag{2}$$

*Under the assumption that $\forall i = 1, \ldots, M \quad d_i \ll \lambda$ , the field inside layers, $\varphi_n$, has no spatial variations. From the continuity of tangential components of the electric field at the boundaries, and also the lack of spatial variations of the field inside each layer, Eqs. (3) and (4) are obtained.*

$$E_{x1} = E_{x2} = E_{x3} = E_{x4} = E_{x5} = E_x \tag{3}$$

$$E_{y1} = E_{y2} = E_{y3} = E_{y4} = E_{y5} = E_y \tag{4}$$

*Using Eq. (2), total $D_x$ can be written in terms of $D_{xi}$ (field inside each layer):*

$$D_x = \frac{d_1 D_{x1} + d_2 D_{x2} + d_3 D_{x3} + d_4 D_{x4} + d_5 D_{x5}}{d_1 + d_2 + d_3 + d_4 + d_5} \tag{5}$$

*By substituting the equation $\forall i = 1, \ldots, 5 \quad D_{x_i} = \varepsilon_i E_{x_i}$ in the equation above, we have:*

$$D_x = \left( \frac{d_1\varepsilon_1 + d_2\varepsilon_2 + d_3\varepsilon_1 + d_4\varepsilon_2 + d_5\varepsilon_1}{d_1 + d_2 + d_3 + d_4 + d_5} \right) E_x \tag{6}$$

*From the similarity of $D_x = \varepsilon_{xx} E_x$ to Eq. (6), an expression for $\varepsilon_{xx}$ is obtained:*

$$\varepsilon_{xx} = \frac{\varepsilon_1 (d_1 + d_3 + d_5) + \varepsilon_2 (d_2 + d_4)}{d_1 + d_2 + d_3 + d_4 + d_5} \tag{7}$$

*In the same manner, $\varepsilon_{yy}$ is obtained and we have:*

$$\varepsilon_{yy} = \varepsilon_{xx} = \varepsilon_{core,\parallel} \tag{8}$$

*From the continuity of the normal component of electric flux density at the boundaries, and also the lack of spatial variations of the field inside each layer, we have:*

$$D_{z1} = D_{z2} = D_{z3} = D_{z4} = D_{z5} = D_z \tag{9}$$

*Using Eq. (2), total $E_z$ can be written in terms of $E_{zi}$ (field inside each layer):*

$$E_z = \frac{d_1E_{z1} + d_2E_{z2} + d_3E_{z3} + d_4E_{z4} + d_5E_{z5}}{d_1 + d_2 + d_3 + d_4 + d_5} \tag{10}$$

*By substituting the equation $\forall i = 1, \dots, 5 \;\; E_{zi} = \frac{D_{zi}}{\varepsilon_i}$ in the equation above, we have:*

$$E_z = \left( \frac{\frac{d_1}{\varepsilon_1} + \frac{d_2}{\varepsilon_2} + \frac{d_3}{\varepsilon_1} + \frac{d_4}{\varepsilon_2} + \frac{d_5}{\varepsilon_1}}{d_1 + d_2 + d_3 + d_4 + d_5} \right) D_z \tag{11}$$

*From the similarity of $E_z = \frac{1}{\varepsilon_{zz}} D_z$ to Eq. (11), an expression for $\varepsilon_{zz}$ is obtained:*

$$\varepsilon_{zz} = \varepsilon_{core,\perp} = \frac{d_1 + d_2 + d_3 + d_4 + d_5}{\frac{1}{\varepsilon_1}(d_1 + d_3 + d_5) + \frac{1}{\varepsilon_2}(d_2 + d_4)} \tag{12}$$

*Finally, the tensor of $\bar{\bar{\varepsilon}}_{core}(\omega)$, which is the permittivity tensor of MQW in the absence of doping, $\bar{\bar{\varepsilon}}_{MQW}(\omega)\big|_{Ne=0}$, is obtained:*

$$\bar{\bar{\varepsilon}}_{core}(\omega) = \begin{bmatrix} \varepsilon_{xx}(\omega) & 0 & 0 \\ 0 & \varepsilon_{yy}(\omega) & 0 \\ 0 & 0 & \varepsilon_{zz}(\omega) \end{bmatrix} \tag{13}$$

*The permittivities of $\varepsilon_1(\omega)$ and $\varepsilon_2(\omega)$ belong to barrier material (AlInAs) and well material (InGaAs), respectively. These materials are dispersive in the mid-infrared band, and in paper,[2] their experimental data have been reported. Notably, here, both materials are undoped.*

## *1.2 Permittivity tensor of doped MQWs*

*In this section, we investigate the case of doped MQWs. The out-of-plane component of permittivity can be written as follows, which contains two resonant terms:[1]*

$$\varepsilon_{MQW,\perp}(\omega) = \varepsilon_{core,\perp}(\omega) + \frac{N_e e^2}{\hbar}\left(\frac{z_{21}^2}{\omega_{21} - \omega + i\gamma_{21}} + \frac{z_{31}^2}{\omega_{31} - \omega + i\gamma_{31}}\right) \tag{14}$$

*Electrons are free in the transverse plane of the semiconductor, and using the Drude model, the in-plane components of permittivity can be written:[1]*

$$\varepsilon_{MQW,\parallel}(\omega) = \varepsilon_{core,\parallel}(\omega) + \frac{\varepsilon_0 \omega_p^2}{-\omega^2 + i\omega\Gamma_D} \tag{15}$$

*In the equation above, $\Gamma_D = \frac{1}{\tau_D}$ is the collision frequency, here, $\tau_D$ is the Drude relaxation time. In the semiconductor compound $In_{0.53}Ga_{0.47}As/Al_{0.48}In_{0.52}As$, in the low-doping case, $\tau_D = 0.1\ [ps]$ is considered.[1]*

*Additionally,* $\omega_p$ *is the plasma frequency, calculated through equation* $\omega_p^2 = \frac{N_e e^2}{\varepsilon_0 m_w^*}$*, here* $m_w^*$ *is the effective mass of electron in the well region. Since most electrons are concentrated in the region of wide well.*

*Finally, the permittivity for the doped MQWs is as follows, which is expressed as a dispersive and anisotropic medium.*

$$\bar{\bar{\varepsilon}}_{MQW}(\omega) = \begin{bmatrix} \varepsilon_{MQW,\parallel}(\omega) & 0 & 0 \\ 0 & \varepsilon_{MQW,\parallel}(\omega) & 0 \\ 0 & 0 & \varepsilon_{MQW,\perp}(\omega) \end{bmatrix} \tag{16}$$